\documentclass[pra,twocolumn, showpacs, superscriptaddress]{revtex4-1}

\usepackage[utf8]{inputenc}
\usepackage{graphicx}
\usepackage{dcolumn}
\usepackage{bbm}
\usepackage{amsmath}
\usepackage{amssymb}
\usepackage{xcolor}

\usepackage{times}
\usepackage{tgtermes}

\renewcommand*{\d}{\ensuremath{\delta}}
\newcommand*{\D}{\ensuremath{\Delta}}
\newcommand*{\s}{\ensuremath{\sigma}}
\newcommand*{\w}{\ensuremath{\omega}}

\newcommand*{\op}[1]{\ensuremath{#1}}
\newcommand*{\hop}[1]{\ensuremath{\op{#1}^{\dagger}}}
\newcommand*{\IdOp}{\ensuremath{\op{\boldsymbol{\mathbbm{1}}}}}
\newcommand*{\bra}[1]{\ensuremath{\langle{#1}\vert}}
\newcommand*{\ket}[1]{\ensuremath{\vert{#1}\rangle}}

\newcommand*{\expect}[1]{\ensuremath{\langle{#1}\rangle}}
\newcommand*{\abs}[1]{\ensuremath{\mathinner{\lvert#1\rvert}}}
\ifdefined\Tr%
	\renewcommand*\Tr{\operatorname{Tr}}
\else%
	\newcommand*\Tr{\operatorname{Tr}}
\fi%

\begin{document}

\title{Dynamic Generation of Topologically Protected Self-Correcting Quantum Memory}

\author{Daniel Becker}
 \email{d.becker@unibas.ch}
 
\affiliation{Department of Physics, University of Basel, Klingelbergstrasse 82, CH-4056 Basel, Switzerland}

\author{Tetsufumi Tanamoto}
\affiliation{Corporate Research and Development Center, Toshiba Corporation, Saiwai-ku, Kawasaki 212-8582, Japan}

\author{Adrian Hutter}
\author{Fabio L. Pedrocchi}
\author{Daniel Loss}

\affiliation{Department of Physics, University of Basel, Klingelbergstrasse 82, CH-4056 Basel, Switzerland}

\date{\today}

\begin{abstract}
We propose a scheme to dynamically realize a quantum memory based on the toric code. The code is generated from qubit systems with typical two-body interactions (Ising, $XY$, Heisenberg) using periodic, NMR-like, pulse sequences. It allows one to encode the logical qubits without measurements and to protect them dynamically against the time evolution of the physical qubits. A weakly coupled cavity mode mediates a long-range attractive interaction between the stabilizer operators of the toric code, thereby suppressing the creation of thermal anyons. This significantly increases the lifetime of the memory compared to the code with noninteracting stabilizers. We investigate how the fidelity, with which the toric code is realized, depends on the period length $T$ of the pulse sequence and the magnitude of possible pulse errors. We derive an optimal period $T_{\text{opt}}$ that maximizes the fidelity.
\end{abstract}


\maketitle

\section{Introduction} \label{sec:Intro}
One of the most promising proposals for the realization of a thermally stable quantum memory is based on topologically ordered phases of matter like Kitaev's toric code \cite{kitaev_fault-tolerant_2003, dennis_topological_2002, bravyi_topological_2010}.

While the toric code allows for topological protection against local imperfections at zero temperature, it is susceptible to thermal fluctuations \cite{nussinov_autocorrelations_2008, castelnovo_entanglement_2007, alicki_statistical_2007, alicki_thermalization_2009}: Anyons that are created at constant energy cost and move freely across the memory without additional energy penalty destroy the stored quantum information in a time that does not increase with the linear size $L$ of the memory. In fact, all two-dimensional (2D) and broad classes of three-dimensional (3D) stabilizer Hamiltonians with local interactions are subject to no-go theorems forbidding stability at finite temperature \cite{bravyi_no-go_2009, haah_logical-operator_2012, yoshida_feasibility_2011}. 3D stabilizer models with local interactions that do not satisfy the criteria of applicability of the no-go theorems have been proposed and studied. None of them, however, is so far expected to be stable at any nonzero temperature \cite{bacon_operator_2006, haah_local_2011, bravyi_energy_2011, michnicki_3-d_2012}.

Despite the intrinsic thermal fragility of the toric code, it is possible to considerably improve the memory lifetime by allowing for long-range repulsive interactions between anyons as proposed in Ref.~\cite{chesi_self-correcting_2010} and later studied in several systems in Refs.~\cite{pedrocchi_quantum_2011,rothlisberger_incoherent_2012,hutter_self-correcting_2012}. In fact, this leads to a suppression of the anyon density in the thermodynamic limit and thus to a memory lifetime increasing polynomially with $L$. Such memories are called \emph{self-correcting}, as their stability against errors caused by the thermal environment is ``built in'' in the sense that no active (measurement-based) error correction is required. In similar approaches, long-range attractive interaction between anyons as a way to suppress their motion across the memory has been proposed in Ref.~\cite{dennis_topological_2002} and later studied in Ref.~\cite{hamma_toric-boson_2009} by coupling the toric code to a bosonic bath. In this case the diffusion of anyons is reduced by the attractive interaction and the memory lifetime increases polynomially with $L$, but the model requires unbounded-strength interactions between anyon operators and the bosonic bath \cite{hamma_toric-boson_2009}. Recently, a three-dimensional model where toric code \emph{stabilizers} are locally coupled to the spins of a ferromagnet has been proposed in Ref.~\cite{pedrocchi_local_2012}. The attractive interaction between stabilizers is then mediated by Goldstone modes and leads to a memory lifetime increasing exponentially with $L$.

In this work, we study a similar model, for which the attractive interaction between stabilizers of the toric code is mediated by cavity modes. This model was first proposed in Ref.~\cite{chesi_self-correcting_2010} and studied in detail in Ref.~\cite{pedrocchi_quantum_2011} as a low-energy effective Hamiltonian of an anisotropic honeycomb model coupled to a cavity mode. In the same sense as for these cavity-based setups, the model proposed here implements a passive error correction which renders it self-correcting against thermal fluctuations. This self-correcting property, however, is limited by pulse imperfections as well as the validity of the perturbation theory used to derive the effective stabilizer interaction.

The toric code Hamiltonian contains many-body interactions that are not directly realized in nature. Nevertheless, several ways have been proposed as to how the toric code (and similar stabilizer codes) can be implemented in practice. Besides the low-energy limit of the honeycomb lattice \cite{kitaev_annals_2006, vidal_perturbative_2008}, it can emerge dynamically as a coarse-grained description of a quantum simulation with discrete time steps using Rydberg atoms \cite{weimer_rydberg_2010}, ions \cite{barreiro_open-system_2011, muller_simulating_2011}, or polar molecules \cite{weimer_quantum_2013, gorshkov_kitaev_2013} in optical traps.  This work is based on another kind of dynamical implementation that employs NMR-like, periodic sequences of short external pulses to induce the dynamics of the code Hamiltonian in solid-state systems, similar to a recent proposal in Ref.~\cite{tanamoto_strategy_2013}. By its use of periodic pulses, this scheme of dynamically generating a desired Hamiltonian dynamics is related to the so-called \emph{dynamical decoupling} methods~\cite{viola_dynamical_1998, *viola_dynamical_1999, zanardi_symmetrizing_1999} that mainly aim to suppress the effect of a given or even unknown system-bath interaction on (time-)average. From an abstract, mathematical point of view, both kinds of pulsing schemes are rather similar insofar as their effect can be captured by taking the time average of Hamiltonian terms or error generators that are transformed by appropriate unitary operations (the pulses). Compared to the dynamical decoupling schemes, the method used here is particularly suited to generate the Hamiltonian dynamics of the planar code with as few linewise applied rotations as possible. Also, in addition to the suppression of decoherence that is caused by a decouplinglike effect of the pulses, our proposal achieves an even stronger stabilization against the thermal environment by a coupling to a nonlocal field (the cavity mode).

Starting from a system of $1/2$ spins coupled to a cavity mode, we show how to generate a toric code with long-range attractive interactions between stabilizers. For perfect pulses, the fidelity $F$ with which the dynamics of the code Hamiltonian is generated only depends on the structure and time period $T$ of the generating sequence reaching the theoretical limit of $F = 1$ for $T \to 0$. In reality, however, unavoidable pulse imperfections require to minimize the number of pulses per time and the fidelity reaches its maximum for a finite period $T_{\text{opt}}$. We derive an explicit formula for $T_{\text{opt}}$ and show that for pulse errors of a relative magnitude around $10^{-4}$ a fidelity of more than $0.99$ can be realized over a long time $t \gtrsim 100 \Delta^{-1}$, where $\Delta$ is the energy scale to create a bare anyon. Finally, we show how to encode a logical qubit by a sequence of magnetic pulses and thus without the need of stabilizer measurements.  

The paper is organized as follows. After the model system is introduced in Sec.~\ref{sec:System}, we show how to generate the dynamics of the toric code Hamiltonian with a periodic pulse sequence in Sec.~\ref{sec:DynamicGeneration}. We study both numerically and analytically how the gate fidelity $F$ of the induced time evolution depends (i) on the structure of the pulse sequence and (ii) on the magnitude of random pulse errors in Sec.~\ref{sec:FidelityLimits}. Furthermore, based on both these limiting factors for $F$, an analytic expression for the optimal sequence period length $T_{\text{opt}} > 0$ is derived that maximizes the fidelity. In Sec.~\ref{sec:StabilizerInteractions}, we derive an effective low-energy Hamiltonian for a pulsed system of qubits that are weakly coupled to a cavity mode. The cavity induces an attractive stabilizer interaction that protects the system against errors caused by a thermal environment. We explain that, with respect to these errors, the resulting memory can be considered as self-correcting. Finally, the measurement-free encoding of logical qubit states into the ground state manifold is explained in Sec. \ref{sec:Preparation}. Appendices \ref{app:WholeCodeOps}--\ref{app:FidelityWithPulseError} contain details about the pulse operations for arbitrarily large qubit arrays and technical derivations regarding the fidelity dependence on the sequence structure and on the magnitude of pulse errors.

\section{The Model System} \label{sec:System}

\begin{figure}

	\vspace*{2mm}%
	\includegraphics[scale=0.8]{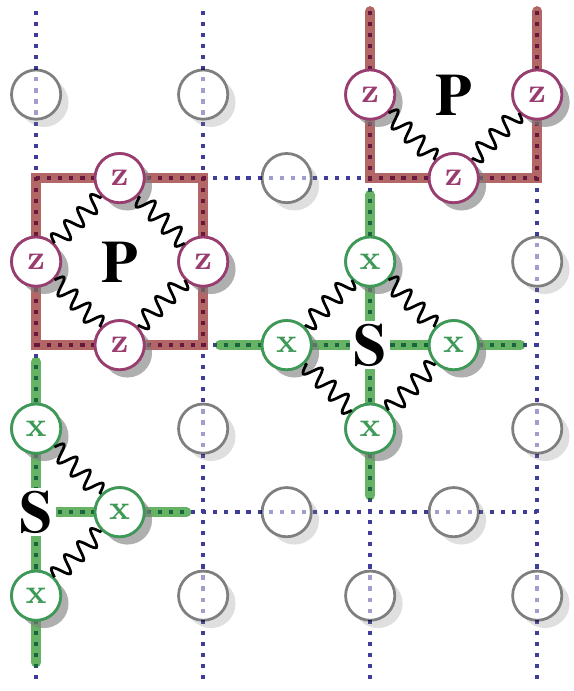}%
	\vskip-\lastskip%
	\caption%
	{%
		\label{fig:ToricHamiltonian}
		(Color online) Schematic of the planar code Hamiltonian equation~\eqref{eqn:ToricHamiltonian} on the square lattice (dotted lines). Plaquettes (P) are associated with the edges of a unit cell and stars (S) with the edges of the cells around a vertex. A circle indicates a Pauli operator acting on a qubit and the wiggly lines connect operators that have to be multiplied to obtain the corresponding stabilizer operator.
	}%

\end{figure}

We consider a quadratic lattice of noninteracting qubits with site-independent level splitting, which are weakly coupled ($0 < \d \ll \D$) to a single cavity mode
\begin{equation}\label{eqn:FreeModel}
	\op{H}_0 (\d) = -4 [\D + \d (\op{b} + \hop{b}) ] \sum_{j} \op{\s}_z^{(j)} + \hbar \w_0 \hop{b} \op{b},
\end{equation}
where $\op{\s}_{\varkappa}^{(j)}$ with $\varkappa \in \{ x, y, z \}$ denotes a Pauli matrix acting on the qubit on lattice site $j$. Operator $\op{b}$ annihilates a photon of energy $\hbar \w_0$ in the cavity. We assume that two-qubit gates can be applied by an external switching on and off of a ``natural'' (system-dependent) two-qubit interaction. For example, in the case of spin qubits realized in single-electron quantum dots an effective nearest-neighbor Heisenberg interaction can be switched electrically by changing the transparency of the tunneling barrier between two dots using gate electrodes~\cite{loss_quantum_1998}. The goal of this work is to show that with the choice of a proper periodic sequence of single-qubit rotations and two-qubit gates, a self-correcting topological quantum memory can be generated dynamically. We show this for the case of the planar code (toric code with boundaries), which is described by Hamiltonian
\begin{equation}\label{eqn:ToricHamiltonian}
		\op{H}_{\text{PC}}
	=	-\D \biggl( \sum_p \op{A}_p + \sum_s \op{A}_s \biggr)
	=:	-\D \sum_a \op{W}_a.
\end{equation}
For a given quadratic lattice of qubits sitting on the edges of a unit cell, indices $p$ and $s$ run over all \emph{plaquettes} and \emph{stars}, respectively. A plaquette is the set $\{p_1, \ldots, p_4 \}$ of qubits sitting on the edge of a single unit cell and the corresponding operator is given by $\op{A}_p = \prod_i \op{\s}_z^{(p_i)}$. Associated to every star is the set $\{s_1, \ldots, s_4 \}$ of qubits around a vertex connecting four neighboring cells, where $\op{A}_s = \prod_i \op{\s}_x^{(s_i)}$. At the edges of the code, the $\op{A}_p$ and $\op{A}_s$ consist of only three Pauli operators. Figure~\ref{fig:ToricHamiltonian} illustrates the structure of $\op{H}_{\text{PC}}$ schematically. The stabilizer operators $\op{W}_a$ are identical to the plaquettes and stars and introduced to obtain a simpler notation. Hence, index $a$ runs through all unit cells and vertices of the quadratic lattice.

We want to stress, however, that the method described here is not restricted to the planar code or the particular free Hamiltonian $\op{H}_0$. In fact, it can in principle be used to generate a quite large class of code Hamiltonians. Restrictions of this method rather concern the practical implementation of a particular scheme in real experiments. Details such as the required fidelities on certain time scales, addressability issues, the accuracy of external operations, etc., have to be examined with respect to a particular physical system. Below we present an example of how to implement the dynamically generated topologically protected quantum memory based on $\op{H}_{\text{PC}}$ in realistic systems.

\section{Dynamic Generation of Planar Code} \label{sec:DynamicGeneration}

In this section, we explain in detail how external pulses and two-qubit operations can be used to dynamically ``generate'' the planar code Hamiltonian from the qubit part of $\op{H}_0 (\d = 0)$. Here and in the following ``generating a Hamiltonian $\op{H}_{\text{av}}$'' is used to mean that, due to external pulses applied between certain discrete (stroboscopic) times, a system with Hamiltonian $\op{H}$ evolves as if its dynamics was governed by $\op{H}_{\text{av}}$---the ``\emph{average Hamiltonian.}'' After a short introduction to average Hamiltonian theory \cite{ernst_principles_1990}, we proceed by showing one possible sequence of pulses that generates $\op{H}_{\text{PC}}$ from $\op{H}_0 (\d = 0)$. Finally, a nonzero coupling of the qubits to a cavity mode [as in Eq. \eqref{eqn:FreeModel}] can be utilized to induce a strongly nonlocal interaction between the anyon operators of $\op{H}_{\text{PC}}$.

\subsection{Average Hamiltonian Theory} \label{subsec:AHTheory}

\begin{figure}

	\vspace*{2mm}%
	\includegraphics[scale=1]{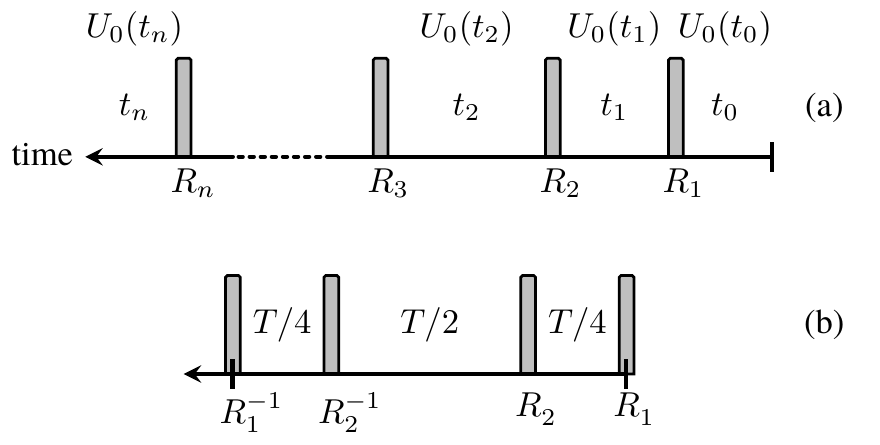}%
	\vskip-\lastskip%
	\caption%
	{%
		\label{fig:GenericPulseSequence}
		(a) Generic pulse (operation) sequence. The unitary operations $\op{R}_i$ (gray columns) with $\prod_{i = 1}^n \op{R}_i = \IdOp$ interrupt periods of (free) propagation with $\op{H}_0$. (b) Structure of a sequence of duration $T$ to generate a plaquette or star operator. A sequence of the same structure is employed to generate (each) one of the quarters of the planar code.
	}%

\end{figure}

Average Hamiltonian theory describes how time-periodic (externally controlled) unitary transformations can be used to let the evolution of one system mimic that of another system of the experimenter's choosing. For the purpose of our paper, it is sufficient to consider the case of $n + 1$ periods $t_0,\ldots,t_n$ of free propagation with $\op{H}_0$, which are separated in time by $n$ unitary transformations $\op{R}_i$ with $i = 1,\ldots,n$ and $\prod_{i = 1}^n \op{R}_i = \IdOp$. It is assumed, that all the $\op{R}_i$ can be performed within a typical time $\tau_i$, where the $\tau_i$ define the smallest time scale of the system.
This is illustrated in Fig. \ref{fig:GenericPulseSequence}(a). The physical time of the entire sequence is then given by $T = t_0 + \sum_{i = 1}^n ( t_i + \tau_i )$, after which the model will have evolved according to the time evolution operator
$
		\op{U}_T
	=	\op{U}_0 (t_n) \op{R}_n \cdots \op{R}_2 \op{U}_0 (t_1) \op{R}_1 \op{U}_0 (t_0)
	\equiv
		\exp \{ -i \tilde{T} \op{H}_{\text{av}} \}
$ with $\tilde{T} =  \sum_{i = 0}^n t_i \lesssim T$.
Here and in the following, $\hbar$ is set to $1$ and we defined $\op{U}_0 (t) = \exp \{ -i t \op{H}_0 \}$. With the Magnus expansion \cite{ernst_principles_1990, magnus_exponential_1954}, such a product of unitary operators can always be written as a single exponential of the average Hamiltonian $\op{H}_{\text{av}}$. Hence, if the pulse sequence is applied periodically, at \emph{integer multiples of $T$} (stroboscopic times) the model system will have evolved as if governed by Hamiltonian $\tilde{T} \op{H}_{\text{av}}/T$. In general, the exact $\op{H}_{\text{av}}$ can only be given as an infinite expansion in orders of $T\D$, where $\D^{-1}$ is (of the order of) the characteristic time scale of the unpulsed system $\op{H}_0$:
\begin{equation}\label{eqn:GeneralMagnusExpansion}
		\op{H}_{\text{av}}
	=
			\op{H}_{\text{av}}^{(0)}
		+	\op{H}_{\text{av}}^{(1)}
		+	\op{H}_{\text{av}}^{(2)} \ldots
\end{equation}
For small enough $T\D$, however, it is often sufficient to only consider the lowest-order term
\begin{equation}\label{eqn:AvH}
		\op{H}_{\text{av}}
	\approx 
		\op{H}_{\text{av}}^{(0)}
	=	\frac{1}{T} \sum_{j = 0}^n \op{H}_j t_j,
\end{equation}
with $\op{H}_j = \hop{\mathcal{R}}_j \op{H}_0 \op{\mathcal{R}}_j$ for $j \ne 0$ and $\op{\mathcal{R}}_j = \prod_{k\le j} \op{R}_k$.

\subsection{Generating Sequence for Planar Code} \label{subsec:GeneratingSequence}

In order for a sequence as shown in Fig. \ref{fig:GenericPulseSequence}(a) to be usable in real experiments and applications it has to be as short as possible and the operations $\op{R}_i$ have to be decomposable into simple elementary operations. We will show how $\op{H}_{\text{PC}}$ can be generated from $\op{H}_0 (\d = 0)$ using only (i) $\pi /2$ single-qubit rotations about the $x$ and $y$ axes and (ii) the two-qubit controlled phase gate $\op{U}_{\text{PG}} = \exp\{ i \pi \op{\s}_z^{(1)} \op{\s}_z^{(2)} / 4 \} = (\IdOp + i\, \op{\s}_z^{(1)} \op{\s}_z^{(2)}) / \sqrt{2}$ between qubits on neighboring sites (denoted by $1$ and $2$). Each of these operations only have to be applicable on rows (columns, diagonals) of qubits simultaneously. In particular, we will not require that physical qubits can be addressed individually, which might be advantageous for certain qubit architectures.

In general, besides single-qubit rotations the generation of $\op{H}_{\text{PC}}$ only requires an operation that allows one to transform Pauli terms $\op{h} = \op{\s}_{\varkappa}^{(p)}$ in the average Hamiltonian according to $\op{\s}_{\varkappa}^{(p)} \rightarrow \op{\s}_{\varkappa'}^{(p)} \op{\s}_{\varkappa''}^{(\bar{p})}$ for at least one axis $\varkappa$. Here, $p = 1, 2$ and $\bar{p} \ne p$ are the site indices of neighboring qubits. The spin axes $\varkappa$, $\varkappa'$, and $\varkappa''$ do not need to be different.

Gate $\op{U}_{\text{PG}}$ is an example of such an operation, which can be seen by considering a sequence of two pulses with $\op{\mathcal{R}}_1 = \op{R}_1 = \op{U}_{\text{PG}}$, and $t_0 = t_2 = 0$ so that $t_1 \equiv T$. According to Eq. \eqref{eqn:AvH}, we obtain the average Hamiltonian to be $\op{h}_{\text{av}} = \hop{U}_{\text{PG}} \op{h} \op{U}_{\text{PG}}$. A straightforward calculation shows that this transformation has nontrivial effects only on terms that contain \emph{exactly one} of either $\op{\s}_x$ or $\op{\s}_y$. Concretely, the transformation converts $\op{\s}_x^{(p)} \rightarrow \op{\s}_y^{(p)} \op{\s}_z^{(\bar{p})}$ and $\op{\s}_y^{(p)} \rightarrow -\op{\s}_x^{(p)} \op{\s}_z^{(\bar{p})}$.

In a system, in which the qubit interaction is Ising-like with $\op{H}_{\text{Ising}} = - J \sum_{\langle jk \rangle} \op{\s}_z^{(j)} \op{\s}_z^{(k)}$ and $J > 0$, where the sum includes all next neighbors $\langle jk \rangle$, the operation $\op{U}_{\text{PG}}$ can be performed by switching the interaction on for a time $\tau = \pi /(4 J)$. Note that similar operations can also be realized based on other nearest-neighbor interactions such as $XY$ or Heisenberg. In the $XY$ case, for example, this role can be played by $\exp \{ \pm i J \tau (\op{\s}_x^{(1)} \op{\s}_x^{(2)} + \op{\s}_y^{(1)} \op{\s}_y^{(2)}) \}$ for $\tau = \pi / (4 J)$, while using a (sightly modified) pulse sequence of similar complexity to the one we present for the Ising case. Analogous operations can be found in the Heisenberg case. Thus, our scheme is not restricted to qubit systems with a particular kind of two-particle interaction.

\begin{figure}

	\vspace*{2mm}%
	\includegraphics[scale=0.8]{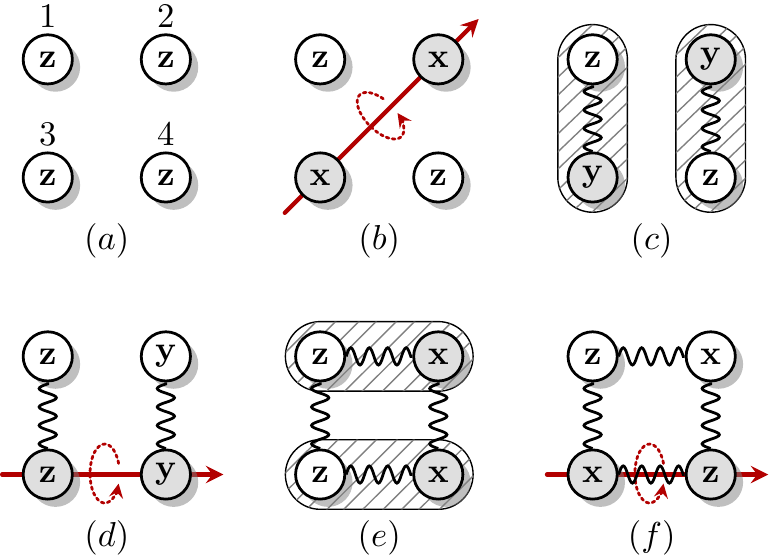}%
	\vskip-\lastskip%
	\caption%
	{%
		\label{fig:FourthOrderTerm}
		(Color online)
		First 6 of 7 elementary steps to generate fourth-order terms in an average Hamiltonian starting from a single-particle Hamiltonian for qubits on a quadratic lattice.
	}%

\end{figure}

We proceed by showing how to generate a plaquette or star operator for a $2 \times 2$ quadratic lattice of free qubits. For simplicity, we consider the case $\D = 1/4$ and $\d = 0$, given by $\op{H}_0 = \sum_{j = 1}^4 \op{\s}_z^{(j)}$, with $j$ denoting the qubits as in Fig. \ref{fig:FourthOrderTerm}(a), and $\op{H}_{\text{av}} = \op{A}_s = \prod_j \op{\s}_x^{(j)}$. The pulse sequence consists of four (complex) operations with $t_0 = t_4 = 0$ and $2 t_1 = 2 t_3 = t_2 = T/2$. Its structure is illustrated in Fig. \ref{fig:GenericPulseSequence} (b), where we have to specify operations $\op{R}_1$ and $\op{R}_2$ so that the sequence yields the desired (lowest-order) average Hamiltonian.

A possible sequence of operations to generate a fourth-order spin Hamiltonian starting from $\op{H}_0$ is illustrated in Fig.~\ref{fig:FourthOrderTerm}, while Table~\ref{tab:IntermediateH} shows the terms of the transformed Hamiltonian for each step. After rotating the qubits along one diagonal [upward in Fig.~\ref{fig:FourthOrderTerm}(b)] about the $y$ axis by $\pi / 2$, second-order terms (indicated by wiggly lines) are generated by applying $\op{U}_{\text{PG}}$ (hatched areas) to two parallel edges of the lattice [vertical in Fig.~\ref{fig:FourthOrderTerm}(c)]. Since applying the entangling gate to the remaining (horizontal) edges in the resulting configuration would yield third- instead of fourth-order terms, the qubits along one of these edges [the lower in Fig.~\ref{fig:FourthOrderTerm}(d)] have to be rotated, so that the $z$ and $y$ terms are interchanged. Subsequent application of $\op{U}_{\text{PG}}$ to the remaining edges [horizontal in Fig.~\ref{fig:FourthOrderTerm}(e)] now generates [amongst others, see Eq. \eqref{eqn:AvHR1}] a fourth-order term.

The last step shown in Fig.~\ref{fig:FourthOrderTerm}(f) is not actually necessary in case of the simple $2 \times 2$ lattice. Depending on whether a plaquette or staright-hand-sideto be generated, one could either $y$ rotate the rightmost or leftmost edge after step (e). In view of the application to a larger planar code like in Fig. \ref{fig:ToricHamiltonian}, however, we interchange the lower edge qubits a second time [Fig.~\ref{fig:FourthOrderTerm}(f)].  A final rotation of the qubits along a properly chosen diagonal then yields the desired fourth-order operator (not shown in the figure). In the case of the planar code on a larger lattice, the additional step is required due to the presence of (third-order) boundary operators in $\op{H}_{\text{PC}}$. It arranges their components in straight lines with those of the interior operators, so that a final rotation along properly chosen diagonals generates a Hamiltonian with only $z$ and $x$ terms, respectively (cf. Appendix \ref{app:WholeCodeOps}).

\begin{table}

	\hbox to \linewidth%
	{%
		\hfill%
		\begin{minipage}{0.8\linewidth}%
			\caption%
			{%
				\label{tab:IntermediateH}
				Hamiltonian terms for each step in Fig.~\ref{fig:FourthOrderTerm} (up to a sign). Pauli operators that changed compared to the previous step are shown against a gray background.
				A final $y$ rotation (not shown) of qubits 1 and 4 yields Eq.~\eqref{eqn:AvHR1}.
			}%
			\begin{ruledtabular}
				\begin{tabular}{lllll}
						(a)
					&	$\op{\s}_z^{(1)}$
					&	$\op{\s}_z^{(2)}$
					&	$\op{\s}_z^{(3)}$
					&	$\op{\s}_z^{(4)}$
					\\[1.5pt]
						(b)
					&	$\op{\s}_z^{(1)}$
					&	\colorbox{gray!35}{$\op{\s}_x^{(2)}$} 
					&	\colorbox{gray!35}{$\op{\s}_x^{(3)}$} 
					&	$\op{\s}_z^{(4)}$
					\\[0pt]
						(c)
					&	$\op{\s}_z^{(1)}$
					&	\colorbox{gray!35}{$\op{\s}_y^{(2)}\;\hspace*{0.45pt} \op{\s}_z^{(4)}$} 
					&	\colorbox{gray!35}{$\op{\s}_z^{(1)}\;\hspace*{0.45pt} \op{\s}_y^{(3)}$} 
					&	$\op{\s}_z^{(4)}$
					\\[0pt]
						(d)
					&	$\op{\s}_z^{(1)}$
					&	$\op{\s}_y^{(2)}$ \colorbox{gray!35}{$\op{\s}_y^{(4)}$} 
					&	$\op{\s}_z^{(1)}$ \colorbox{gray!35}{$\op{\s}_z^{(3)}$} 
					&	\colorbox{gray!35}{$\op{\s}_y^{(4)}$} 
					\\[1.5pt]
						(e)
					&	$\op{\s}_z^{(1)}$
					&	\colorbox{gray!35}{$\op{\s}_z^{(1)} \op{\s}_x^{(2)}\;\hspace*{0.45pt} \op{\s}_z^{(3)} \op{\s}_x^{(4)}$} 
					&	$\op{\s}_z^{(1)} \op{\s}_z^{(3)}$
					&	\colorbox{gray!35}{$\op{\s}_z^{(3)} \op{\s}_x^{(4)}$} 
					\\[0pt]
						(f)
					&	$\op{\s}_z^{(1)}$
					&	$\op{\s}_z^{(1)} \op{\s}_x^{(2)}$ \colorbox{gray!35}{$\op{\s}_x^{(3)} \op{\s}_z^{(4)}$} 
					&	$\op{\s}_z^{(1)}$ \colorbox{gray!35}{$\op{\s}_x^{(3)}$} 
					&	\colorbox{gray!35}{$\op{\s}_x^{(3)} \op{\s}_z^{(4)}$} 
				\end{tabular}
			\end{ruledtabular}
		\end{minipage}
		\hfill%
	}%

\end{table}

The candidate for $\op{R}_1$ to generate a star operator from $\op{H}_0$ is therefore the product of all the operations of steps (b)--(f) plus a $y$ rotation of qubits 1 and 4 with
\begin{equation}\label{eqn:FourthOrderOp}
		\op{R}_1
	=	\op{U}_{x \leftrightarrow z}^{(2,3)}\,
		\op{U}_{\text{PG}}^{(1,3)} \op{U}_{\text{PG}}^{(2,4)}\,
		\op{U}_{y \leftrightarrow z}^{(3,4)}\,
		\op{U}_{\text{PG}}^{(1,2)} \op{U}_{\text{PG}}^{(3,4)}\,
		\op{U}_{x \leftrightarrow z}^{(3,4)}\,
		\op{U}_{x \leftrightarrow z}^{(1,4)},
\end{equation}
where $\op{U}_{\text{PG}}^{(jk)}$ entangles qubits $j$ and $k$ and $\op{U}_{x \leftrightarrow z}$ denotes a $\pi/2$ rotation about the axis perpendicular to $x$ and $z$. Note that $\op{R}_1$ acts \emph{on the system} by performing the elementary steps \emph{in reverse order}. If we have only $\op{R}_1$ or, equivalently, set $\op{R}_2 = \IdOp$, we obtain average Hamiltonian
\begin{equation}\label{eqn:AvHR1}
		\op{H}_1
	=		\op{\s}_x^{(1)} \op{\s}_x^{(2)} \op{\s}_x^{(3)} \op{\s}_x^{(4)}
		+	\op{\s}_x^{(1)} \op{\s}_x^{(3)}
		+	\op{\s}_x^{(3)} \op{\s}_x^{(4)}
		-	\op{\s}_x^{(1)}.
\end{equation}
Apart from the desired star operator, it contains unwanted single- and two-particle by-products (see columns 1,3, and 4 in Table~\ref{tab:IntermediateH}). These can be removed from the average Hamiltonian with an $\op{R}_2$ that yields a Hamiltonian $\op{H}_2 = \op{R}_1^{-1} \op{R}_2^{-1} \op{H}_0 \op{R}_2 \op{R}_1$ [see Eq.~\eqref{eqn:AvH}], whose first- and second-order terms have the opposite sign compared to $\op{H}_1$. Note that if such an operation $\op{R}_2$ were considered as just another step of the sequence in Fig. \ref{fig:FourthOrderTerm}, it would have to be performed \emph{prior to (b)}: With $\op{H}_0' := \op{R}_2^{-1} \op{H}_0 \op{R}_2$, we can write $\op{H}_2 = \op{R}_1^{-1} \op{H}_0' \op{R}_1$. Hence, we need to find an operation $\op{R}_2$ that changes $\op{H}_0$ to $\op{H}_0'$ in such a way that the \emph{subsequent} application of $\op{R}_1$ results in the desired sign flips.

It turns out that the simple ansatz to just rotate qubits 1 and 3 by an angle of $\pi$ about either the $y$ or $z$ axis (in an attempt to add one sign to each first- and second-order term, while adding two signs to the fourth-order operator) will not do the trick. A straightforward calculation shows that qubit 3 has to be rotated as well to give the correct result. Thus, with $\op{R}_2 = \op{U}_y^{(1,3,4)} (\pi)$ and using $\op{H}_3 = \op{H}_1$ we obtain the average Hamiltonian $\op{H}_{\text{av}} = (\op{H}_1 + \op{H}_2) / 2 = \op{\s}_x^{(1)} \op{\s}_x^{(2)} \op{\s}_x^{(3)} \op{\s}_x^{(4)}$. Here, operator $\op{U}_{\varkappa}^{(i_1,\ldots,i_n)} (\phi)$ rotates qubits $i_1, \ldots, i_n$ about axis $\varkappa = x,y,z$ by angle $\phi$. Finally, $\op{R}_2$ has to be decomposed into operations that are applied linewise. One possible way to do this is
\begin{equation}\label{eqn:DecomposeR2}
		\op{R}_2
	=	\op{U}_{y}^{(1,4)} (\pi/2) \, \op{U}_x^{(3,4)} (\pi) \, \op{U}_{y}^{(1,4)} (\pi/2).
\end{equation}

\begin{figure}

	\vspace*{2mm}%
	\includegraphics[scale=1]{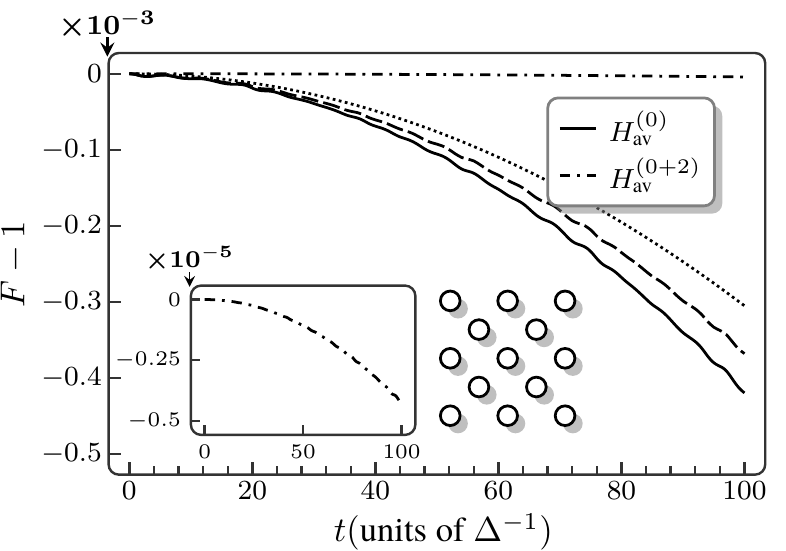}%
	\vskip-\lastskip%
	\caption%
	{%
		\label{fig:GFidelity}
		Numerical gate fidelity versus time of a dynamically generated planar code of $N = 13$ qubits (inset) coupled to a cavity mode with three dimensional Fock space. For $\D = \w_0$, $\d/\D = 0.1$, and $T = \D^{-1}/8$, already the lowest-order average Hamiltonian yields $F > 99.9 \%$ for times $t \le 100 \D^{-1}$ (solid line). Adding the second-order contribution decreases the deviation from perfect fidelity by another two orders of magnitude (dashed and dotted line, plot in inset). Based on the two leading orders of the average Hamiltonian, the exact fidelity can be approximated reasonably well, both numerically (dashed line) and analytically (dotted line).
	}%

\end{figure}

With a pulse sequence of the exact same structure and similar (generalized) $\op{R}_1$ and $\op{R}_2$, we can also generate either multiple plaquette or star operators on an arbitrary large quadratic lattice. Details about both operations are given in Appendix \ref{app:WholeCodeOps}. We point out that only one-quarter (even or odd subsets of plaquettes or stars) of $\op{H}_{\text{PC}}$ can be generated with one such sequence. This is due to the fact that only disjoint, i.e., half of the pairs of qubits, can be coupled in an entangling step per lattice dimension, to generate fourth- and third-order terms of the correct structure. Since each quarter of the code is generated for one-quarter of the time only, the single-particle energy of Hamiltonian $\op{H}_0 (\d = 0)$ from Eq. \eqref{eqn:FreeModel} was set to be $-4 \D$ to yield a $\op{H}_{\text{PC}}$ with energy gap $-\D$.

\section{Fidelity Limits due to Sequence Structure and Pulse Errors} \label{sec:FidelityLimits}

A suitable measure for the similarity of the dynamically generated dynamics with that of the toric code Hamiltonian is given by the gate fidelity
\begin{equation}\label{eqn:GateFidelity}
	F(t) = \abs{\Tr [ \exp(i t \op{H}_{\text{av}}) \op{U}_{\text{P}} (t) ]}/2^N,
\end{equation}
where $\op{U}_{\text{P}} (t)$ denotes the evolution operator of the pulsed system. The gate fidelity indicates how well an arbitrary basis of the whole Hilbert space evolves according to $\op{H}_{\text{av}}$. By contrast, a state-specific fidelity only quantifies how well the dynamics is described by a particular $\op{H}_{\text{av}}$ for that same specific set of states. For example, a high value for a fidelity that measures how much of an encoded state remains in the ground state manifold over time, is consistent with the dynamics of any Hamiltonian that coincides with that of $\op{H}_{\text{av}}$ on the ground state manifold. For the generation of the stabilizer interaction by means of a cavity, however, it is important that besides the ground states all excited states propagate according to $\op{H}_{\text{av}}$, as well. Hence, in contrast to state fidelities, only a gate fidelity $F(t) \approx 1$ for $\op{H}_{\text{av}} = \op{H}_{\text{PC}}$ can be used as an indicator that the cavity indeed induces the stabilizer interaction.

For decreasing sequence duration of the order of or shorter than $\D^{-1}$, by simply generating each four quarters alternatingly with a period of $T/4$, the lowest-order average Hamiltonian
\begin{equation}\label{eqn:ZerothOrderHAv}
		\op{H}_{\text{av}}^{(0)}
	=		[1 + \d/\D (\op{b} + \hop{b}) ] \op{H}_{\text{PC}}
		+	\w_0 \hop{b} \op{b}
\end{equation}
describes the system dynamics with increasing fidelity over longer times even between the end points of the sequence. With a suitably symmetrized version (cf. Fig.~\ref{fig:FullSymmetricSequence} in Appendix~\ref{app:SecondOrderCorrection}), all odd-order contributions to $\op{H}_{\text{av}}$ [see Eq.~\eqref{eqn:GeneralMagnusExpansion}] can be completely suppressed, so that the leading-order deviation is given by
\begin{equation}\label{eqn:MainDeviation}
	\begin{split}
				\op{H}_{\text{av}}^{(2)}
		&	=	\frac{\w_0 T^2 \d}{8 \D} \mathbf{Q}
				[%
						\mathcal{M} \mathbf{Q}^{\dagger}
					+	\mathbf{V} (\op{b} + \hop{b})
				] \quad\text{with} \\
				\mathcal{M}
		&	=	\frac{\d}{3 \D}
				\begin{pmatrix}
					-6 & -5 & -5 & -5 \\
					-5 &  0 &  1 &  1 \\
					-5 &  1 &  6 &  7 \\
					-5 &  1 &  7 & 12
				\end{pmatrix}
		\quad
				\mathbf{V}
			=	\frac{\w_0}{4}
				\begin{pmatrix}
					 7 \\
					 1 \\
					-3 \\
					-5
				\end{pmatrix}\!,
	\end{split}
\end{equation}
where $\mathbf{Q} = (\op{Q}_1, \ldots, \op{Q}_4)$ is a vector consisting of the four quarters $\op{Q}_i = -\D \sum_{\{a_i\}} \op{W}_{a_i}$ of the planar code defined by appropriate, disjoint sets of anyon indices $\{a_i\}$.

Figure~\ref{fig:GFidelity} shows the time-dependent gate fidelity for perfect pulses and a lattice of $N = 13$ qubits (inset) coupled to a cavity with three-dimensional Fock space $\{\ket{n\pm1}, \ket{n}\}$, which is evaluated with a numerically exact Chebyshev expansion of the time evolution operator \cite{tal-ezer_accurate_1984}. The system parameters are $T = \D^{-1}/8$, $\d = 0.1 \D$, and $\w_0 = \D$. Already for $\op{H}_{\text{av}} = \op{H}_{\text{av}}^{(0)}$ (solid line), the fidelity does not drop below $99.9 \%$ before $t = 100 \D^{-1}$, while adding the second-order contribution from Eq.~\eqref{eqn:MainDeviation} decreases the deviation from the perfect fidelity by another two orders of magnitude (dashed and dotted line, plot in inset). In reality, however, the fidelity will be lower than this theoretical maximum due to errors in the pulsing scheme, decoherence by the noisy environment, and fluctuations of the microwave beam that is used to induce anyon interactions. Compared to systems that directly realize the toric code Hamiltonian (rather than just its dynamics), e.g., in some low-energy limit of some suitable local lattice Hamiltonian, these effects will decrease the lifetime of codewords. Nevertheless, the numerical results indicate that $\op{H}_{\text{av}}^{(0)} + \op{H}_{\text{av}}^{(2)}$ describes the perfect-pulse dynamics of the system very well and therefore allows one to estimate the size of deviations from the intended Hamiltonian \eqref{eqn:ZerothOrderHAv} that arise solely by the structure of the pulse sequence. This is illustrated by the dashed line in Fig.~\ref{fig:GFidelity}, which gives
\begin{equation}\label{eqn:DefF2}
	F^{(2)} := \abs{\Tr [ \exp(i \op{H}_{\text{av}}^{(0)} t) \exp(-i \{ \op{H}_{\text{av}}^{(0)} + \op{H}_{\text{av}}^{(2)} \} t) ]} / 2^N
\end{equation}
as result of a numerically exact calculation for the 13-qubit system. Its quadratic behavior is very similar to that of the pulsed system, while the quantitative deviations are due to terms of order 4 and higher. In Appendix \ref{app:SecondOrderCorrection} it is shown that $F^{(2)}$ can be approximated by $F^{(2)} \approx 1 - c_{\text{av}} t^2$ with
\begin{equation}\label{eqn:ApproxF2}
		c_{\text{av}}
	=	\frac{1}{3} \Bigl( \frac{5 \w_0 L(L-1) \d\,\D}{16} \Bigr)^2 T^4
\end{equation}
for planar codes of arbitrary linear size $L$, which we define as the length of the larger quadratic sublattice for codes arranged as in Figs.~\ref{fig:ToricHamiltonian}, \ref{fig:GFidelity}, and \ref{fig:planarR}. For example, the code with $13 = 3^2 + 2^2$ qubits has length $L = 3$, while the code with $25 = 4^2 + 3^2$ qubits has length $L = 4$. For $L=3$, the approximate $F^{(2)}$ is given by the dotted line in Fig.~\ref{fig:GFidelity}. Although Eq.~\eqref{eqn:ApproxF2} systematically overestimates the fidelity of the pulsed system $F$, it is sufficiently accurate to provide the bounds $1 < (1 - F)/(c_{\text{av}} t^2) < 2$ for deviations due to higher-order corrections to $\op{H}_{\text{av}}^{(0)}$, as long as $c_{\text{av}} t^2 \ll 1$.

\begin{figure}
	\vspace*{2mm}%
	\begin{center}
		\includegraphics[scale=1]{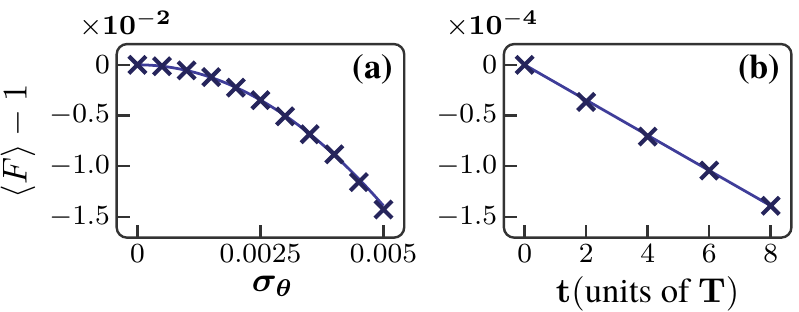}
	\end{center}
	\vskip-\lastskip%
	\caption%
	{%
		\label{fig:FGateWithError}
		(Color online)
		Average of numerical gate fidelity $\expect{F}$ with imperfect pulses and $T = \D^{-1}/8$. (a) shows $\expect{F} - 1$ versus the width $\s_{\theta}$ of the Gaussian distributed pulse errors for $t = \D^{-1}$, while (b) shows the fidelity versus time for $\s_{\theta} = 5 \times 10^{-4}$. The solid lines are evaluated using Eq.~\eqref{eqn:ApproxFErr} and agree well with the numerical data (crosses).
	}
\end{figure}

While the deviations that are due to the structure of the pulse sequence decrease rapidly for smaller periods $T$, errors caused by imperfect pulses will generally increase with the pulsing rate. As a consequence, a finite magnitude of pulse errors entails an optimal, finite value of $T$, for which the total deviations due to both error sources are minimal. We estimate the effect of pulse imperfections using a simple model with errors manifesting in small random deviations $\d\theta \ll 1$ from the intended angle $\theta_0$ in $\op{\tilde{R}} (\d\theta) = \exp(i [\theta_0 + \d\theta] \op{S} / 2 )$, where $\op{S} = \op{\s}_{\varkappa}^{(j)}$ for rotations about axis $\varkappa$ and $\op{S} = \op{\s}_z^{(j)}\op{\s}_z^{(j + 1)}$ for the phase gate, respectively. The error distribution with width $\s_{\theta} = \sqrt{\expect{\d\theta^2}} > 0$ is assumed to be Gaussian, equal for all considered kinds of pulses, and unbiased ($\expect{\d\theta} = 0$). In Appendix \ref{app:FidelityWithPulseError}, we show that this error model leads to an average gate fidelity $\expect{F} \approx 1 - c_{\text{err}} t - \alpha c_{\text{av}} t^2$ with $1 < \alpha < 2$ and
\begin{equation}\label{eqn:ApproxFErr}
		c_{\text{err}}
	=	( 18 L^2 - 10 L + 7 ) \frac{\s_{\theta}^2}{2T}
\end{equation}
for $c_{\text{err}} t \ll 1$. As shown in Fig.~\ref{fig:FGateWithError}, this simple approximation (solid lines) agrees well with the numerical results (crosses) of the $L = 3$ code for $T = \D^{-1}/8$ and pulse errors $\lesssim 10^{-2}$. The fidelity exhibits a linear and quadratic behavior in dependence on the time and on the magnitude of pulse errors, respectively. Thus, by minimizing the total deviations $c_{\text{err}} t + \alpha c_{\text{av}} t^2$ with respect to $T$, we can estimate the optimal period length to be
\begin{equation}\label{eqn:TOpt}
		T_{\text{opt}}
	=	\biggl( \frac{96 [2L (9L - 5) + 7] \s_{\theta}^2}{[5 L (L-1) \d \w_0 \D ]^2 \alpha t } \biggr)^{1/5}.
\end{equation}
For example, to maximize $\expect{F}$ for the $L = 3$ code with $\w_0 = \D$, $\d = 0.1 \D$, and an error magnitude of $\s_{\theta} = 0.001$ after a propagation time of $t = 100 \D^{-1}$, we have to set $T_{\text{opt}} \approx 0.1 \D^{-1}$, which results in a fidelity of about $91 \%$. For $\s_{\theta} = 10^{-4}$, we obtain $T_{\text{opt}} \approx 0.04 \D^{-1}$ and $\expect{F} (100 \D^{-1}) \gtrsim 99\%$. When realizing the optimal sequence period, the total deviation from perfect fidelity scales with the linear size of the memory like $O(L^{12/5})$. This limits the gain in stability that is achievable by the induced stabilizer interaction (see Sec.~\ref{sec:StabilizerInteractions}). In an experimental realization this leads to an optimal $L$ that maximizes the lifetime of the memory.

\section{Suppression of Thermal Anyons due to Cavity-Induced Stabilizer Interactions}\label{sec:StabilizerInteractions}

In this section, we derive the effective long-range attractive interaction between stabilizers that is induced by the coupling of a cavity mode to the entire lattice of physical qubits. As long as the coupling is weak ($\d \ll \w_0, \D $), a Schrieffer-Wolff transformation \cite{bravyi_schriefferwolff_2011} yields the effective low-energy Hamiltonian for a cavity that is kept in a Fock state $\ket{n}$ with a constant number $n$ of photons. Aside from small deviations due to $\op{H}_{\text{av}}^{(2)}$ and pulse imperfections, the total average Hamiltonian is then given by
\begin{equation}\label{eqn:HEffQubit}
		\op{H}
	=	-	\D \sum_{a} \op{W}_a
		-	\frac{\d^2}{\w_0} \sum_{a\ne a'} \op{W}_{a} \op{W}_{a'}.
\end{equation}
The interaction term, which describes an effective long-range interaction between the stabilizer operators, couples every pair of stabilizers of the code via absorption and emission of a single photon. Note that this term in the effective Hamiltonian~\eqref{eqn:HEffQubit} is at least two orders of magnitude stronger than the one that appears in the second-order average Hamiltonian in Eq.~\eqref{eqn:MainDeviation} for typical values of $T_{\text{opt}}$ and $\d \ll \w_0 \approx \D$.

It was shown in Ref.~\cite{pedrocchi_quantum_2011} that the energy penalty for the creation of anyons grows with $L^2$ for a code Hamiltonian~\eqref{eqn:HEffQubit} that features an attractive constant long-range interaction between stabilizers. As a consequence, the lifetime of logical qubit states (or codewords) due to \emph{independent single-qubit errors} that are created by the \emph{thermal environment} increases exponentially with $L^2$ or, equivalently, with the number of stabilizer operators $W_{a}$. In a system that is a direct realization of Hamiltonian~\eqref{eqn:HEffQubit}, the density of anyons therefore goes to zero in the thermodynamic limit and without the need for any measurements-based, active error-correction procedure, the memory retains its information indefinitely. In other words, the coupling to the cavity mode allows one to passively protect the quantum information against thermal fluctuations. In this sense the memory is called self-correcting.

Note that, while being an element of repeated external manipulation, the pulsing of the system does not in itself constitute an active error-correction procedure: Neither does it require one to extract information about the system state at any point in time (by measuring some system properties) in order to detect an error, nor to manipulate the system in a way that is conditional on the result of (such) an error syndrome measurement. Rather, any error-correcting effect is purely passive and achieved by employing a time invariant sequence of pulses to create an environment that hinders anyon creation. Consequently, we can regard the time invariant pulsing as an integral part of the system---a system that is passively protected against thermal fluctuations and therefore self-correcting in the sense explained above. In contrast to a direct realization of Hamiltonian~\eqref{eqn:HEffQubit}, however, the dynamical quantum memory has a finite lifetime in the thermodynamic limit. Besides the factors discussed in Sect.~\ref{sec:FidelityLimits}, it is limited by (i) fluctuations in the cavity mode, (ii) a breakdown of the perturbation theory used to derive the effective interaction in Eq.~\eqref{eqn:HEffQubit} for large $L$, and (iii) system-bath interaction processes with coherence times that are of the order of or shorter than $T$.

In summary, the following constraints between energies and time scales have to be fulfilled by every implementation of the dynamically generated quantum memory:
\begin{equation}\label{eqn:Constraints}
	\d \ll \D, \w_0,\quad
	T \ll T_1,T_2,\quad
	\beta^{-1} \ll \D,
\end{equation}
where $\beta^{-1}$ is the thermal energy and $T_1$ and $T_2$ are the relaxation and dephasing times of the physical qubits, respectively. The latter provide time scales both for the rates of errors created in the code as well as for the process of error creation itself. Further restrictions may appear depending on the details of the physical realization.

\section{Preparation of Codeword States} \label{sec:Preparation}

In addition to generating the dynamics of the planar code Hamiltonian, the pulsing method can also be used to prepare codewords $\{\ket{\bar{0}}, \ket{\bar{1}}\}$ without the need to perform measurements on the system \cite{tanamoto_strategy_2013}. Typically, a codeword of the planar code is prepared by consecutive projective measurements of all (commuting) star operators performed on initial state $\ket{0,0,\ldots,0}$ that is obtained by preparing all physical qubits in the spin-up state and which is already a simultaneous eigenstate of all the plaquette operators with eigenvalue(s) $+1$.

Specifically, a codeword in the ground state multiplet of $\op{H}_{\text{PC}}$---one that is free of anyons---is obtained by
\begin{equation}\label{eqn:StandardPreparation}
		\ket{\bar{0}}
	=	\prod_s \frac{1}{\sqrt{2}} (\IdOp + \op{A}_s) \ket{0,\ldots,0}.
\end{equation}
The nonunitary projection operators $\propto (\IdOp + \op{A}_s)$ cannot, in general, be implemented with external pulses that are essentially unitary operations. However, if for one of the qubits $s_k \in \{s_1,\ldots,s_4\}$ of star $A_s$, the state prior to the projection can be written as $\ket{\psi_i} = \ket{0}_{s_k} \otimes \ket{\phi}$, where $\ket{\phi}$ is an arbitrary state in the joint Hilbert space of all $N - 1$ remaining qubits, then, by dynamically generating $-\D\op{\tilde{A}}_s (k)$ with $\op{\tilde{A}}_s (k) = -\op{\s}_y^{(s_k)} \prod_{l \ne k} \op{\s}_x^{(s_l)}$ for a time $t = \pi/(4\D)$, we get
\begin{equation}\label{eqn:SingleProjector}
		\exp [i \pi \op{\tilde{A}}_s (k) /4 ] \ket{\psi_i}
	=	\frac{1}{\sqrt{2}} (\IdOp + \op{A}_s) \ket{\psi_i},
\end{equation}
which can be traced back to the identity $-i \op{\s}_y \ket{0} = \op{\s}_x \ket{0}$.

\begin{figure}
	\vspace*{2mm}%
	\begin{center}
		\includegraphics[scale=0.8]{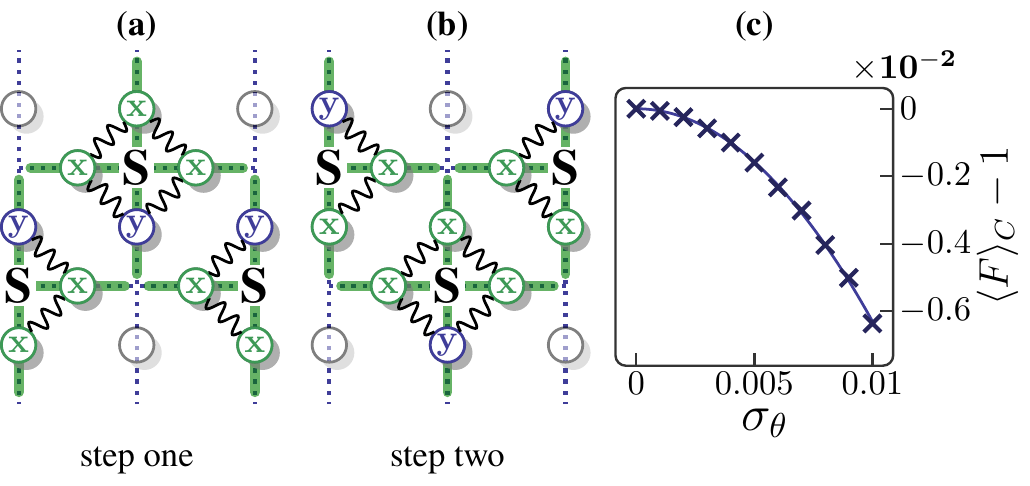}
	\end{center}
	\vskip-\lastskip%
	\caption%
	{%
		\label{fig:PrepLogicalZero}
		(Color online)
		(a) and (b): Two steps for preparing $\ket{\bar{0}}$ from initial state $\ket{0,\ldots,0}$ for the $L=3$ code. The modified star operators are obtained by rotating one $\op{\s}_x$ to $-\op{\s}_y$. In each of the steps, the indicated modified stabilizers are generated for a (effective) time $t = \pi / (4 \D)$ and the $\op{\s}_y$ are chosen such that none of the stabilizers from step one acted on $y$-qubits from step two. (c) Average codeword fidelity $\expect{F}_C$ as a function of pulse errors. Numerics (crosses) agree well with the analytical result (solid line).
	}
\end{figure}

As per Eqs.~\eqref{eqn:StandardPreparation} and \eqref{eqn:SingleProjector}, one $\op{\tilde{A}}_s (k)$ for each star operatoright-hand-sideto be generated, to prepare state $\ket{\bar{0}}$. The order of their application and the set of rotated qubit terms has to be chosen such that for each $\op{\tilde{A}}_s (k)$, none of the previously applied $\op{\tilde{A}}_{s'} (k')$ has acted on the qubit $s_k$. For the $L = 3$ code, this is a two-step process and schematically illustrated in Fig. \ref{fig:PrepLogicalZero}. In each step, one-half of the modified star operators are generated. While there are $4 \times 3 \times 3$ equivalent possibilities to choose three spin operators to rotate from $x$ to $y$ in step one, the corresponding operators in step two are uniquely determined. This procedure works particularly well for the $L = 3$ code, as it mainly consists of edge operators; for $L > 3$, the preparation requires a larger number of steps. It is clear, however, that such a set of steps can always be found and generated with linewise rotations and entangling gates only: In the worst case, the procedure consists of $L (L-1)$ steps---one per star operator.

Figure~\ref{fig:PrepLogicalZero}(c) shows the codeword fidelity, i.e., the probability $F_C := \abs{\bra{\bar{0}} \op{U}_{\text{prep}} \ket{0,\ldots, 0}}^2$ to find the system in state $\ket{\bar{0}}$ after the preparation sequence $\op{U}_{\text{prep}}$, as a function of the pulse errors and with $\d = 0$ for the time of preparation \footnote{The preparation also works for nonzero coupling although with considerably lower fidelity. For example, if $\w_0 \approx \D$ and $\d = 0.1 \D$, we obtain $F_C \approx 90 \%$.}. Again the fidelity decreases quadratically with $\s_{\theta}$ and agrees well with the simple approximation from Appendix \ref{app:FidelityWithPulseError}: Counting the number of mutually inverse pulses in the preparation sequence corresponding to Figs.~\ref{fig:PrepLogicalZero}(a) and \ref{fig:PrepLogicalZero}(b) yields $F_C \approx 1 - 63 \s_{\theta}^2$ for the $L = 3$ code.

\section{Conclusions} \label{sec:Conclusions}

We showed how to dynamically realize a quantum memory based on noninteracting qubits that is stabilized against thermal fluctuations by a weakly coupled microwave cavity. Properly designed, periodic sequences of pulses, implementing single-qubit rotations and controlled phase gates, can be used to induce the time evolution of Kitaev's toric code Hamiltonian. This allows one to prepare logical qubit states without stabilizer measurements and to protect them both against local sources of decoherence and thermal fluctuations for times much longer than the longest time scale of the free qubit system, even when pulse imperfections are taken into account. Furthermore this method is fairly versatile, as it can be generalized to qubit systems with Ising, $XY$, and Heisenberg interactions as well as to arbitrary (2D stabilizer) codes based on local stabilizers in a straightforward manner. We provided simple, accurate analytical estimates for the gate and codeword fidelities of our method as a function of the system parameters, the period length $T$ of the pulse sequence, and the magnitude of randomly distributed pulse errors. This allows us to estimate the maximum tolerable size of pulse errors and to optimize tunable system parameters, such as $T$ or the coupling of the qubits to the cavity $\d$, in order to maximize the lifetime of codeword states.

\begin{acknowledgments}

We would like to thank Andreas Nunnenkamp for discussions. This work was financially supported by the Swiss SNF, the NCCR Nanoscale Science, the NCCR Quantum Science and Technology, and IARPA.

\end{acknowledgments}


\appendix

\section{Operations to Generate the Planar Code} \label{app:WholeCodeOps}

\begin{figure*}
	\vspace*{2mm}%
	\centerline%
	{%
		\includegraphics[scale=0.5]{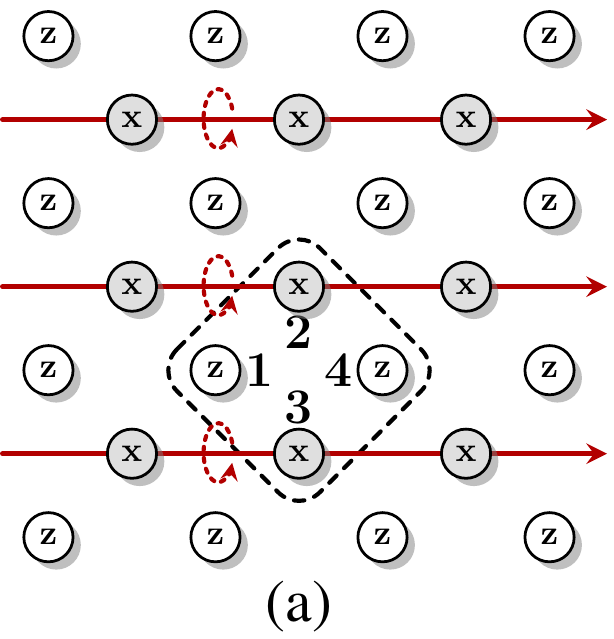}
		\hspace*{2mm}
		\includegraphics[scale=0.5]{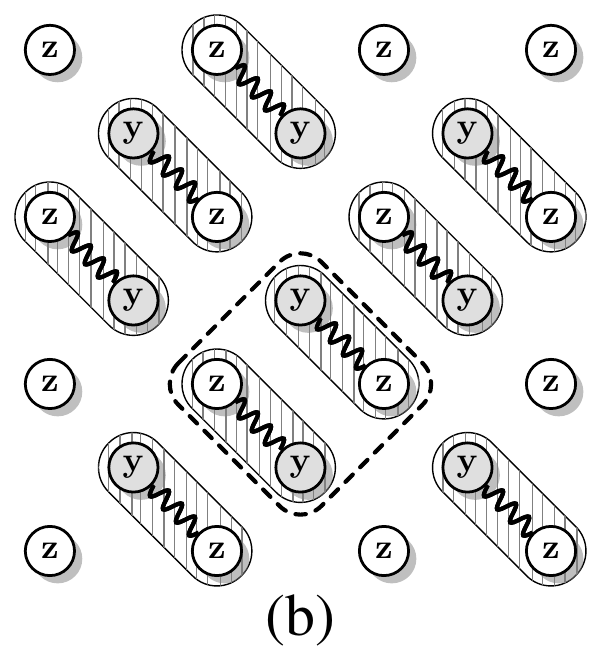}
		\hspace*{2mm}
		\includegraphics[scale=0.5]{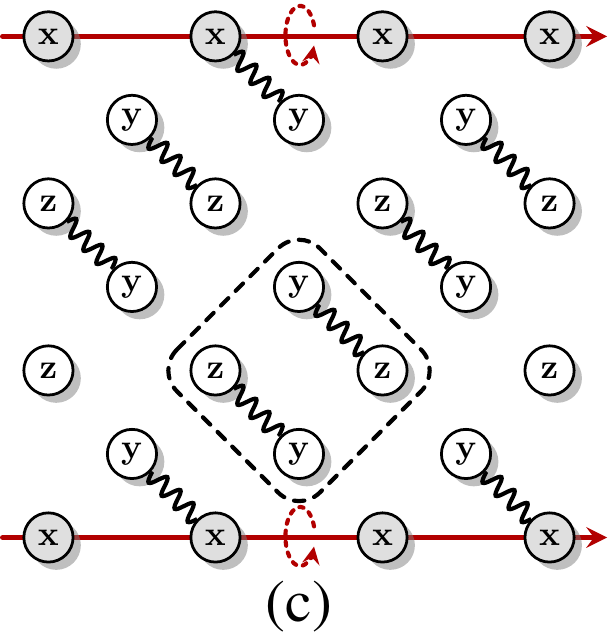}
		\hspace*{2mm}
		\includegraphics[scale=0.5]{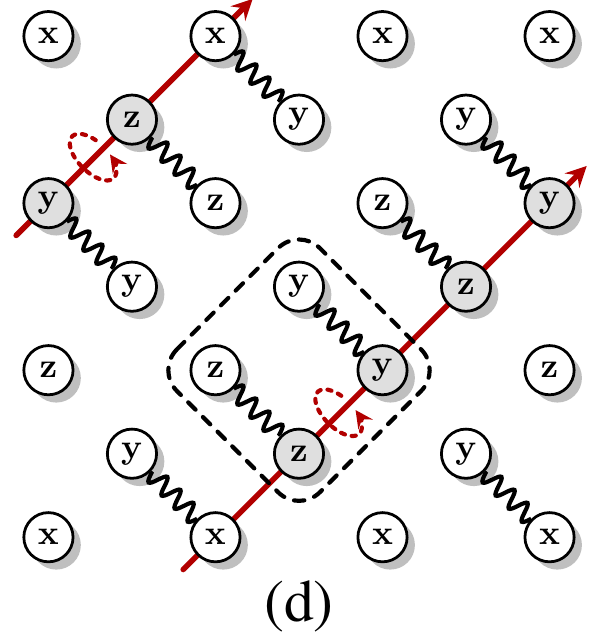}
	}%
	\vspace*{2mm}
	\centerline%
	{%
		\includegraphics[scale=0.5]{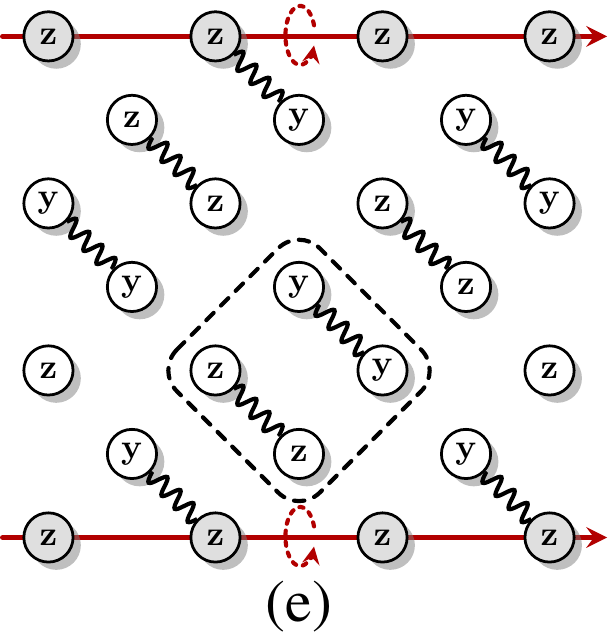}
		\hspace*{2mm}
		\includegraphics[scale=0.5]{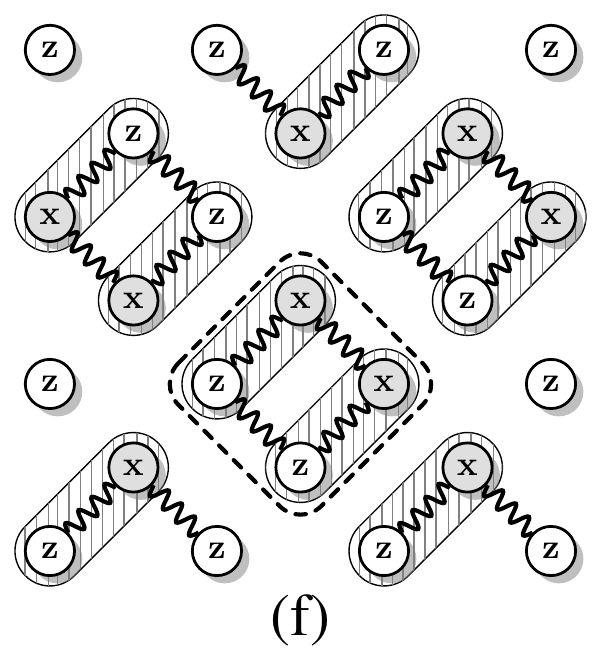}
		\hspace*{2mm}
		\includegraphics[scale=0.5]{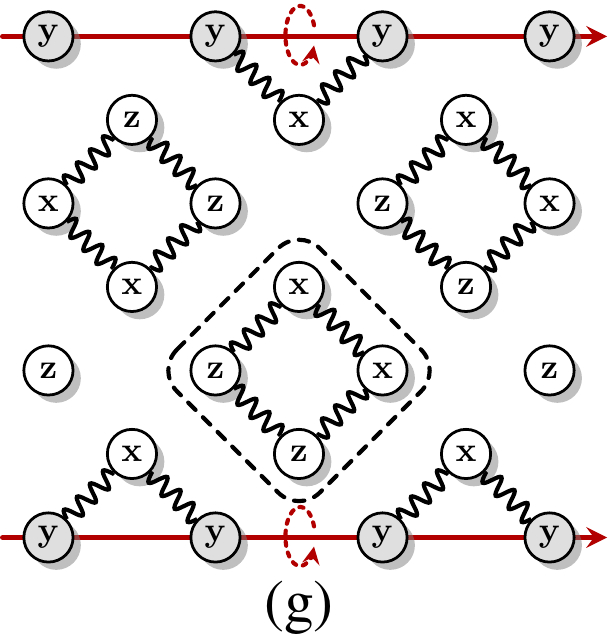}
		\hspace*{2mm}
		\includegraphics[scale=0.5]{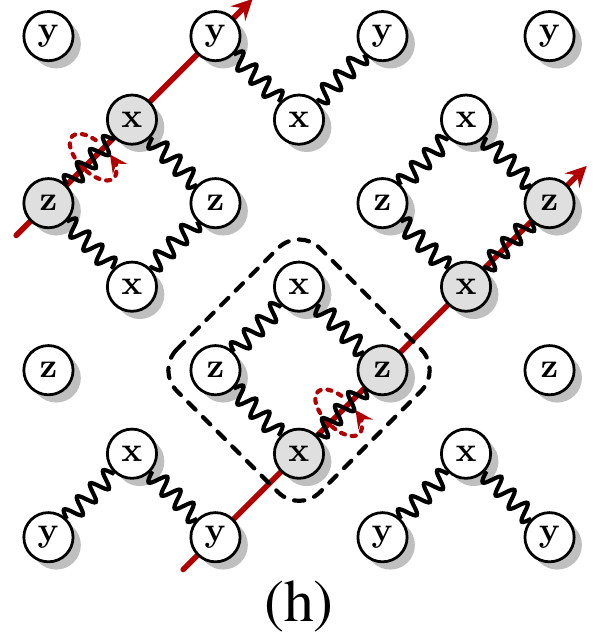}
	}%
	\centerline{\line(1,0){250}}%
	\vspace*{4mm}%
	\centerline%
	{%
		\includegraphics[scale=0.5]{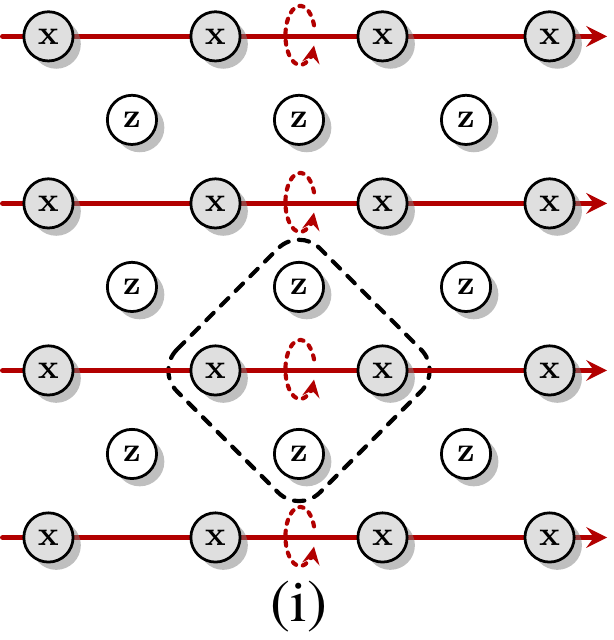}
		\hspace*{2mm}
		\includegraphics[scale=0.5]{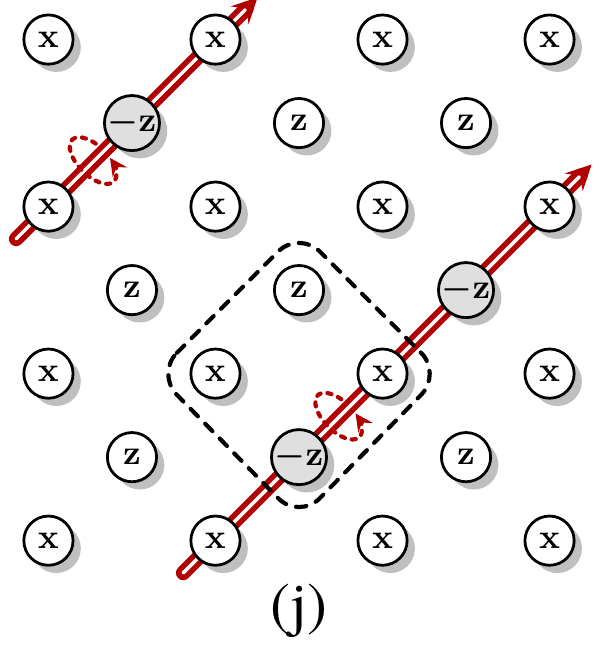}
		\hspace*{2mm}
		\includegraphics[scale=0.5]{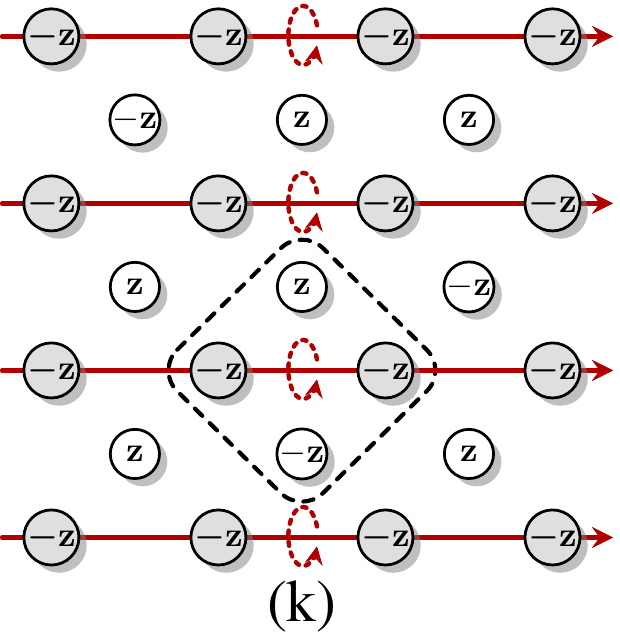}
	}%
	\vskip-\lastskip%
	\caption%
	{%
		\label{fig:planarR}
		(Color online)
		(a)--(h) Schematic illustration of the first eight of nine steps of $\op{R}_1$ to generate a quarter of the planar code Hamiltonian for a lattice of $25$ qubits. When restricted to an interior unit cell (one is marked by a dashed rectangle) these steps reduce to those shown in Fig. \ref{fig:FourthOrderTerm}. (i)--(k) Decomposition of $\op{R}_2$ into a sequence of linewise rotations. Single and double arrows represent rotations by $\pi/2$ and $\pi$, respectively.
	}
	
\end{figure*}

We present a particular choice of operations $\op{R}_1$ and $\op{R}_2$ that can be used to dynamically generate (one-quarter of) the planar code when applied to $\op{H}_0$ as part of the sequence shown in Fig. \ref{fig:GenericPulseSequence}(b). They are a generalization of the operations given in Sec.~\ref{sec:DynamicGeneration} to a quadratic qubit lattice of arbitrary size, i.e., their restriction to an interior unit cell equals Eqs. \eqref{eqn:AvHR1} and \eqref{eqn:DecomposeR2}. The additional operations compared to the $2 \times 2$ lattice are needed to generate proper (third-order) boundary terms, while using strictly linewise rotations and entangling operations. Schematic illustrations of these operations are provided in Fig.~\ref{fig:planarR} for a finite lattice of $25$ qubits. 

The generalized $\op{R}_1$ consists of nine steps, eight of which are shown in Figs. \ref{fig:planarR}(a)--(h). Note that compared to Fig. \ref{fig:FourthOrderTerm}, the lattice is rotated by $45^{\circ}$. After even rows of qubit terms are transformed from $z$ to $x$ by a $\pi/2$ rotation about the $y$ axis, disjoint pairs of qubits are entangled along one of the two sets of parallel diagonals [downward in Fig.~\ref{fig:planarR}(b)] so that pairs on different diagonals are aligned along the perpendicular direction. The following three steps [Fig.~\ref{fig:planarR}(b)--\ref{fig:planarR}(e)] interchange one of the $z$ and $y$ terms along the perpendicular direction for every interior cell, while leaving the qubits that belong to boundary \emph{terms} unchanged. In that context, it is important to notice, that one of the three qubits contributing to a boundary term does not actually lie on that boundary. The rotations that interchange $y$ and $z$ terms [Fig.~\ref{fig:planarR}(d)] are performed about the $x$ axis. Thus, by rotating the terms on the upper and lower boundaries to $\op{\s}_x$ [Fig.~\ref{fig:planarR}(c)] \emph{prior} to the interchange operation, their relative orientation is protected. After the interchange, the boundary qubits are rotated back [Fig.~\ref{fig:planarR}(e)]. The diagonals, along which qubits are interchanged in step $4$ are chosen in the following way: (i) they are perpendicular to the entangling bonds between the second-order terms, (ii) only one pair of qubits is interchanged per unit cell, and (iii) no $y$ qubit belonging to a boundary term is affected. This choice is always possible. The remaining pairs of qubits are entangled in step $6$ [Fig.~\ref{fig:planarR}(f)] to form third- and fourth-order terms. Finally, the qubits of those interior cells that were interchanged in step $4$, have to be changed back [subfigure (h)] in step $8$, again after protecting the relative orientation of the boundaries by a proper rotation [Fig.~\ref{fig:planarR}(g)]. The last step $9$ (not shown) depends on whether plaquette or star operators are about to be generated. All qubits on the same \emph{row} have the same orientation after step $8$. The axis, about which each row has to be rotated in step $9$, is therefore uniquely determined.

Operation $\op{R}_2$ eliminates unwanted single- and two-particle terms in the average Hamiltonian. For the $\op{R}_1$ given above, it has to flip the signs of all qubits (i) on odd rows and (ii) on the intersections of even rows with those diagonals, whose $z$ and $y$ orientations were interchanged in step $4$ of $\op{R}_1$. This operation can be decomposed into three steps, as illustrated in Figs.~\ref{fig:planarR}(i)--(k). First, odd rows are rotated from $z$ to $x$ [Fig.~\ref{fig:planarR}(i)]. A subsequent $\pi$ rotation [double arrows in Fig.~\ref{fig:planarR}(j)] about the $x$ axis of all qubits along the diagonals of step $4$ of $\op{R}_1$ then flips the sign of qubits (ii). Finally, the same operation as in step $1$ is performed once more on the qubits on odd rows [Fig.~\ref{fig:planarR}(k)].

The remaining three-quarters of the code Hamiltonian are readily given by $\pi/2$ rotations of all those steps about the center of the code. Although the choice of $\op{R}_1$ and $\op{R}_2$ is not unique and might by modified depending on a particular experimental realization, the complexity of the operations cannot be decreased by more than one or two steps. A simplification can be achieved, e.g., by choosing entangling operations and rotations that conform to the fourfold symmetry of the quadratic lattice. Since the resulting $\op{R}_1$ and $\op{R}_2$ are not as easily generalized to lattices of arbitrary size and not as easily applicable in a linewise fashion, we do not consider them here. 

\section{Second-Order Corrections to Average Hamiltonian} \label{app:SecondOrderCorrection}

\begin{figure}
	\vspace*{2mm}%
	\begin{center}
		\includegraphics[scale=1]{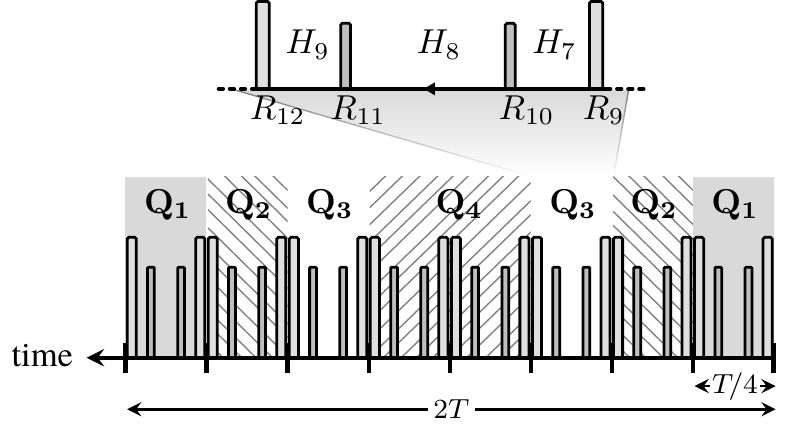}
	\end{center}
	\vskip-\lastskip%
	\caption%
	{%
		\label{fig:FullSymmetricSequence}
		Symmetric pulse sequence for the generation of the planar code. The order in which the quarters $\op{Q}_i$ are generated is reversed after every period $T$. With respect to $T$, the 24 toggling frame Hamiltonians $\op{H}_i = \hop{\mathcal{R}}_i \op{H}_0 \op{\mathcal{R}}_i$ are symmetric, while the $32$ operations $\op{R}_j$ are antisymmetric.
	}
\end{figure}

The exact average Hamiltonian $\op{H}_{\text{av}} = \sum_{n=0}^{\infty} \op{H}_{\text{av}}^{(n)}$ is given by an infinite expansion in orders of $T\D$ and can, in principle, be evaluated to arbitrary orders using the Magnus expansion. For $T \ll \D^{-1}$ this expansion converges quickly and only the lowest orders are relevant to describe the dynamics of the pulsed system. Furthermore, all odd-order contributions can be suppressed by symmetrizing the generating pulse sequence in time with respect to its center \cite{ernst_principles_1990}. A symmetric sequence for the planar code is obtained by reversing the order in which the different quarters are generated after every period $T$, as illustrated in Fig. \ref{fig:FullSymmetricSequence}. The corresponding Hamiltonians $\op{\mathcal{Q}}_i$ with $i = 1, \ldots, 4$ are given by
\begin{equation}\label{eqn:QuarterH}
	\begin{split}
				\op{\mathcal{Q}}_i
		&	=	[1 + \d/\D (\op{b} + \hop{b}) ] \op{Q}_i
				+	\w_0 \hop{b} \op{b}.
	\end{split}
\end{equation}
Taken as a whole, the resulting sequence consists of $32$ operations, which give rise to $24$ toggling frame Hamiltonians. With respect to time $T$, these Hamiltonians are mirror symmetric ($\op{H}_i = \op{H}_{25 - i}$ for $13 \le i \le 24$), while the sequence of $32$ operations $\op{R}_j$ is antisymmetric ($\op{R}_j = \hop{R}_{33 - j}$ for $17 \le j \le 32$). With this, the leading-order deviation from Eq.~\eqref{eqn:ZerothOrderHAv} is essentially given by $\op{H}_{\text{av}}^{(2)} \propto (T\D)^2$. However, as the sub-sequences to generate the (average Hamiltonians) $\op{\mathcal{Q}}_i$ are symmetric as well, the corresponding second-order deviations scale as $(T\D/4)^2$ and can be neglected for the calculation of $\op{H}_{\text{av}}^{(2)}$. Hence, based on a sequence with effective toggling frame Hamiltonians $\op{\mathcal{Q}}_i$, we obtain
\begin{equation}\label{eqn:HigherOrderHAv}
		\op{H}_{\text{av}}^{(2)}
	=	\frac{T^2}{384} \hspace*{-1.75mm} \sum_{1 \le j < k \le l \le 8} \hspace*{-2mm} (1-\d_{kl}/2) [\op{\mathcal{Q}}_l, [\op{\mathcal{Q}}_k, \op{\mathcal{Q}}_j]],
\end{equation}
where $\d_{kl}$ is the Kronecker delta and $\op{\mathcal{Q}}_{j>4} = \op{\mathcal{Q}}_{9 - j}$. Note that since the corresponding spin Hamiltonians $\op{Q}_j$ commute, all higher-order contributions to $\op{H}_{\text{av}}$ vanish identically for $\d = 0$ in which case Eq.~\eqref{eqn:ZerothOrderHAv} gives the exact average Hamiltonian. In the case of a finite coupling to the cavity, $\op{H}_{\text{av}}^{(2)}$ evaluates to Eq.~\eqref{eqn:MainDeviation} using
\begin{equation}\label{eqn:}
	\begin{split}
			[\op{\mathcal{Q}}_l, [\op{\mathcal{Q}}_k, \op{\mathcal{Q}}_j]]
		=	- \d \w_0
			[
					2 \d \op{Q}_l
				+	\w_0(\op{b} + \hop{b})
			]
			(\op{Q}_k - \op{Q}_j).
	\end{split}
\end{equation}

To derive an approximation for the decrease in the gate fidelity $F^{(2)}$ [see Eq.~\eqref{eqn:DefF2}] that is caused by the leading-order correction $\op{H}_{\text{av}}^{(2)}$, we transform the propagator $\exp [ -i t (\op{H}_{\text{av}}^{(0)} + \op{H}_{\text{av}}^{(2)}) ]$ into the interaction picture with $\op{H}_{\text{av}}^{(0)}$ and $\op{H}_{\text{av}}^{(2)}$ playing the role of the noninteracting system and the interaction, respectively. The transformation of the latter can be evaluated analytically yielding
\begin{equation}\label{eqn:Hav2IntPic}
		\op{H}_{\text{av}}^{(2)} (t)
	=	\op{A} + \cos (\w_0 t) \op{B} + \sin(\w_0 t) \op{C}
\end{equation}
with Hermitian operators
\begin{equation}\label{eqn:Hav2IntPicABC}
	\begin{split}
				\op{A}
		&	=	\frac{\d^2\w_0 T^2}{48 \D^2}
				\mathbf{Q}
				\begin{pmatrix}
					 9 & -7 & -19 & -25 \\
					11 &  3 &  -7 & -13 \\
					11 &  5 &   3 &  -1 \\
					11 &  5 &   5 &   9
				\end{pmatrix}\!
				\mathbf{Q}^{\dagger}, \\
				\op{B}
		&	=	\frac{\d \w_0 T^2}{8 \D} \mathbf{Q} \mathbf{V}
				\Bigl[ (\hop{b} + \op{b}) - \frac{2 \d}{\w_0 \D} \sum_j \op{Q}_j \Bigr], \\
				\op{C}
		&	=	\frac{i \d \w_0 T^2}{8 \D} \mathbf{Q} \mathbf{V} (\hop{b} - \op{b}).
	\end{split}
\end{equation}
To determine the approximate fidelity for times $t \gg \D^{-1}$, we replace the time-dependent second-order contribution by its average over the cavity's oscillation period $T_C = 2\pi / \w_0$. It is given by $\op{\bar{H}}_{\text{av}}^{(2)} := T_C^{-1} \int_0^{T_C} \op{H}_{\text{av}}^{(2)} (t)\,d t = \op{A}$. To second-order in $\op{A}$, the fidelity $F^{(2)}$ can then be written as
\begin{equation}\label{eqn:F2Expand}
		F^{(2)}
	\approx
		\abs{\Tr_{\text{QB}} \Bigl( \IdOp - i t \op{A} - \frac{t^2}{2} \op{A}^2 \Bigr) }/ 2^N,
\end{equation}
where $\Tr_{\text{QB}}$ denotes the trace over the qubit subsystem. Using $\Tr_{\text{QB}} (\op{Q}_k \op{Q}_l) = \d_{kl} L (L-1) / 2 $ (the number of anyon operators per code quarter for $k=l$), a straightforward calculation yields Eq.~\eqref{eqn:ApproxF2}.

\section{Gate Fidelity in Presence of Pulse Errors} \label{app:FidelityWithPulseError}

To second-order in $\d\theta$, an imperfect pulse $\op{\tilde{R}} (\d\theta) = \exp(i [\theta_0 + \d\theta] \op{S} / 2 )$ can be approximated by
\begin{equation}\label{eqn:ErrorDeviation}
		\op{\tilde{R}} (\d\theta)
	\approx
			\Bigl[
					\Bigl( 1 - \frac{\d\theta^2}{8} \Bigr) \IdOp
				+	i \frac{\d\theta}{2} \op{S}
			\Bigr] \op{R},
\end{equation}
where $\op{R} \equiv \op{\tilde{R}} (0)$ is the perfect pulse operator. To estimate the statistical effect of these random errors on the fidelity of the whole planar code sequence, it is sufficient to consider the case of two inverse operations with \emph{independent} errors flanking a period of free propagation:
\begin{equation}\label{eqn:TwoErrorPulses}
	\begin{split}
			&	\hop{\tilde{R}} (\d\theta_2) \op{U}_0(t) \op{\tilde{R}} (\d\theta_1) \\
		=	&	\hop{\tilde{R}} (\d\theta_2) \op{\tilde{R}} (\d\theta_1) \hop{\tilde{R}} (\d\theta_1) \op{U}_0(t) \op{\tilde{R}} (\d\theta_1) \\
		=	&	\hop{\tilde{R}} (\d\theta_2) \op{\tilde{R}} (\d\theta_1) \exp [-i t \hop{\tilde{R}} (\d\theta_1) \op{H}_0 \op{\tilde{R}} (\d\theta_1) ]
	\end{split}
\end{equation}
The third term on the right-hand-side can be interpreted as a time evolution operator with effective Hamiltonian $\op{H}_1 + \d\op{H}_1$, where $\op{H}_1 = \hop{\tilde{R}}(0) \op{H}_0 \op{\tilde{R}}(0)$ is the perfect-pulse toggling frame Hamiltonian and
\begin{equation}\label{eqn:ErrorTFH}
		\d\op{H}_1
	=	-	\frac{\d\theta^2}{4}(\op{H}_1 - \op{S}\op{H}_1\op{S})
		+	i \frac{\d\theta}{2} [\op{S}, \op{H}_1].
\end{equation}
In the case of $\op{H}_0 = -\D \op{\s}_x^{(1)}$, $\op{S} = \op{\s}_y^{(1)}$, and $\theta_0 = \pi/2$, for example, application of $\op{\tilde{R}} (\d\theta)$ leads to an average effective Hamiltonian $-\D[(1 - \d\theta^2/2)\op{\s}_z^{(1)} + \d\theta \op{\s}_x^{(1)}]$. The remaining terms from the right-hand-side of Eq.~\eqref{eqn:ErrorTFH} gives
\begin{equation}\label{eqn:ErrorOpChain}
	\begin{split}
			&	\hop{\tilde{R}} (\d\theta_2) \op{\tilde{R}} (\d\theta_1) \\
		=	&	\Bigl[
							\Bigl( 1 - \frac{\d\theta_2^2}{8} \Bigr) \IdOp
						-	i \frac{\d\theta_2}{2} \op{S}
				\Bigr]
				\Bigl[
							\Bigl( 1 - \frac{\d\theta_1^2}{8} \Bigr) \IdOp
						+	i \frac{\d\theta_1}{2} \op{S}
				\Bigr] \\
		=	&		\Bigl( 1 - \frac{\d\theta_1^2 + 2 \d\theta_1 \d\theta_2 + \d\theta_2^2}{8} \Bigr) \IdOp
				+	i \frac{\d\theta_1 - \d\theta_2}{2} \op{S}.
	\end{split}
\end{equation}
For the estimate of how much the statistical expectation value of the gate fidelity as a function of $\d\theta$ deviates from the perfect-pulse value, we only need to consider terms that are proportional to the identity operator (as the trace of all Pauli operators vanishes) and $\d\theta_i^2$ (as for independent, unbiased errors $\expect{\d\theta_1 \d\theta_2} = 0$). The only such terms are contained in the first summand on the right-hand-side of Eq.~\eqref{eqn:ErrorOpChain} and yield $\expect{F} \approx 1 - \s_{\theta}^2/4$ for a single pair of mutually inverse imperfect operations. Consequently, for a nested sequence of $n$ such operations, the deviation amounts to $n$ times that value. Hence, by counting the number $n(L)$ of mutually inverse operations in the generating sequence for a planar code of length $L$, we obtain $\expect{F} \approx 1 - t\, n(L) \s_{\theta}^2/(4T)$ for stroboscopic times $t = m T$ with integer $m$. Based on the operations shown in Appendix \ref{app:WholeCodeOps}, the number $n(L)$ for the whole generating sequence evaluates to $n(L) = 36 L^2 - 20 L + 14$ and we obtain $\expect{F} \approx 1 - c_{\text{err}} t - c_{\text{av}} t^2$ with $c_{\text{err}}$ given by Eq.~\eqref{eqn:ApproxFErr}.


\bibliographystyle{apsrev}



\end{document}